\documentclass[11pt,aps,amsmath,latexsym,amsfonts]{article}
\usepackage{jheppub}

\usepackage{graphicx}
\usepackage{epsfig}
\usepackage{amsfonts,amssymb,amsmath}
\usepackage[hang]{subfigure}
\usepackage{epstopdf}
\usepackage{xcolor}
\pdfoutput=1

\usepackage[a4paper,top=2in, bottom=0in, left=2in, right=0in]{geometry}

\usepackage{color}
\let\normalcolor\relax

\newcommand{\be}{\begin{equation}}
\newcommand{\ee}{\end{equation}}
\newcommand{\ba}{\begin{eqnarray}}
\newcommand{\ea}{\end{eqnarray}}

\title{Critical Phenomena of Charged de Sitter Black Holes in Cavities}
\author{Fil Simovic$^{1,2}$, Robert B. Mann$^2$}
\affiliation[1]{Perimeter Institute for Theoretical Physics,\\
	31 Caroline St. N.,Waterloo, Ontario N2L 2Y5, Canada}
\affiliation[2]{Department of Physics and Astronomy, University of Waterloo,\\
	Waterloo, Ontario N2L 3G1, Canada}
\emailAdd{fil.simovic@gmail.com, rbmann@uwaterloo.ca}
\abstract{We examine the thermodynamic behaviour of four-dimensional charged and uncharged de Sitter black holes enclosed in an isothermal cavity, in the extended phase space where the cosmological constant is treated as a thermodynamic pressure. We demonstrate the presence of a novel pressure-dependent phase transition 
in a compact region of phase space
that does not appear in asymptotically anti-de Sitter black holes,  and find a highly non-linear equation of state that does not lead to the usual interpretation of a van der Waals fluid.}
\keywords{de Sitter black holes, cavity, criticality and phase transitions}

\begin{document}
	\maketitle

\section{Introduction}

The study of black hole thermodynamics has proved to be an incredibly fruitful path towards our understanding of not only classical gravity, but quantum gravity, information theory, thermodynamics, and quantum field theory on curved spacetime as well. The remarkable observation that black holes can be assigned thermodynamic properties such as entropy and temperature, and that these quantities obey the laws of thermodynamics, has led to many insights into the microscopic structure of gravity, whose exact nature remains poorly understood.
\\

Of particular interest is the study of black hole phase transitions, the first example of which was discovered by Hawking and Page in 1983 \cite{HawkingPage1983}. That black holes can undergo phase transitions just like ordinary thermodynamic systems is interesting in its own right, but also provides us with a powerful new tool to study strongly coupled systems like quark-gluon plasmas and superfluid systems through the AdS/CFT correspondence \cite{Maldacena1998} \cite{Policastro2002}. The formulation of the AdS/CFT correspondence, which relates a $(d+1)$-dimensional gravitational theory in anti-de Sitter space to a $d$-dimensional conformal field theory on the boundary of AdS, has motivated a broad range of investigations into asymptotically AdS black holes, where a wealth of interesting phenomena occur.  One subject of particular interest has been the thermodynamic phase structure of these black holes, which we discuss in more depth below.
\\

More recently, the formulation of the de Sitter/conformal field theory correspondence (dS/CFT) \cite{Strominger2001}, as well as the physical relevance of de Sitter black holes in cosmology, has motivated the study of the thermodynamics of asymptotically de Sitter black holes \cite{Sekiwa2006}.  This subject is not well understood, primarily because new
issues arise when dealing with  the presence of the cosmological horizon. The existence of two horizons precludes the assignment of a single equilibrium temperature to the entire spacetime, requiring us to either adopt an {\it effective temperature} approach \cite{Urano2009} or consider each horizon as a separate thermodynamic system, as in \cite{Sekiwa2006}. Furthermore, isolated de Sitter black holes evaporate due to Hawking radiation, unlike AdS black holes where reflecting boundary conditions at infinity ensure thermal stability. 
\\

To overcome these issues, one approach is to consider de Sitter black holes that are enclosed in an isothermal cavity at fixed temperature. This allows us to define a grand canonical ensemble in which the cavity acts as a reservoir, and allows for thermodynamically stable black holes to exist. Such an approach was first explored by Brown \cite{Brown1990}, who demonstrated the stability of such an ensemble, and later Carlip and Vaidya \cite{CarlipVaidya2003}, who found a Hawking-Page-like phase transition in both the asymptotically flat and de Sitter cases. 
\\

In recent years, the idea that the cosmological constant can act as a thermodynamic pressure has been extensively studied \cite{Kastor2009}. In this {\it extended phase space}, the cosmological constant acts like a pressure according to the identification\footnote{In $d$ dimensions the relationship is $P=-\dfrac{\Lambda}{8\pi}=-\textrm{sgn}(\Lambda)\dfrac{(d-1)(d-2)}{16\pi l^2}$, where $l$ is the AdS (or dS) length scale} $\Lambda=-8\pi P$. The variation of $\Lambda$ requires the presence of a pressure-volume term in the first law, which leads to novel thermodynamic behavior. In de Sitter space, the cosmological constant is positive, so the variable $P$ is better thought of as a {\it tension} rather than a pressure. For $\Lambda < 0$, these extended phase space systems have been found to have a wealth of interesting new features such as triple points, reentrant phase transitions, and the emergence of polymer-like and superfluid-like phase structures, providing a wealth of new phenomena to study \cite{Kubiznak2016}. In particular, much attention has been given to the striking analogy these systems make with the classical van der Waals fluid \cite{Kubiznak2012}.
\\
 
The purpose of this paper is to explore the thermodynamics of de Sitter black holes in a cavity in this extended phase space. To this end, we consider both neutral and charged de Sitter black holes in an isothermal cavity. We find that de Sitter black holes in a cavity exhibit behaviour analogous to that of their AdS counterparts, with a Hawking-Page phase transition appearing for uncharged black holes, and a small/large black hole phase transition appearing in the charged case. Despite this similarity, we also find that the equation of state for these black holes has a non-linear temperature dependence and does not exhibit behaviour characteristic of a van der Waals fluid. Finally, we find a new type of {\it compact} first-order phase transition in charged de Sitter black holes, which exists within a finite pressure range and is unlike any phase transition seen in asymptotically AdS black holes.
\\

In Section 1 we discuss our general procedure for studying the critical phenomena of de Sitter black holes. In Section 2 we examine in detail the uncharged case, finding that black holes undergo a standard Hawking-Page phase transition despite the inclusion of the pressure-volume term in the first law. In Section 3 we examine charged de Sitter black holes, finding a number of new phenomena, including a small to large black hole phase transition analogous to the kind found in AdS spacetimes, and a pressure-dependent compact phase transition.

\section{Thermodynamics of Black Holes in a Cavity}

Our treatment of black hole thermodynamics follows the methods used by Braden, Brown, Whiting, and York \cite{Brown1990}, and generalizes the recent work of Carlip and Vaidya \cite{CarlipVaidya2003}. We work in the grand canonical ensemble, where the temperature and electric potential are specified at a boundary whose radius is fixed. In de Sitter spacetime, the back reaction of the radiation near the horizon will tend to lower the temperature of the spacetime \cite{Parikh2002}. We ignore these effects in the present analysis, assuming that the equilibration timescale is small compared to the typical timescale of the phase transitions in the system. Unlike previous work in this area, we will also interpret the cosmological constant as a thermodynamic pressure according to the relation 
\begin{equation}
P=-\frac{\Lambda}{8\pi}
\end{equation}
where $P<0$ because $\Lambda$ is positive.  As noted previously, although this makes $P$ a tension, we shall continue to refer to it as pressure. This will allow us to study the equation of state of these black holes in the extended phase space, where the analogy between the first law of thermodynamics and black hole thermodynamics is complete \cite{Kubiznak2016}.
\\

Our program is as follows. We first evaluate the Einstein-Maxwell action with appropriate boundary terms and constraints imposed. From this action we derive the relevant thermodynamic quantities in this particular ensemble, namely the mass $M$, entropy $S$, temperature $T$, and energy $E$. Then, working in the extended phase space where the cosmological constant is identified as a thermodynamic pressure ($\Lambda=-8\pi P$), we enforce the first law of thermodynamics and the Smarr relation, and from this determine what the thermodynamic volume $V$ and surface tension $\lambda$ of the system are. With these quantities, we can define the Helmholtz free energy $F=E-TS$ and determine where phase transitions occur in the system.  Finally, we consider the equation of state $P(T,V)$ for these systems and briefly comment on the extent to which these phase transitions are analogous to the van der Waals phase transitions that occur in everyday fluids.

\subsection{The Action and Thermodynamic Quantities}

The derivation of the various thermodynamic quantities of interest begins with the Euclidean action for the metric $g_{\mu\nu}$ over a region $\mathcal{M}$ with boundary $\partial \mathcal{M}$ and gauge field $A_{\mu}$, which is given by the Einstein-Hilbert action with the Gibbons-Hawking boundary term:

\begin{equation}
I=-\frac{1}{16\pi}\int_{\mathcal{M}}\!\!d^4x\sqrt{g}\,\big(R-2\Lambda+F^2\big)+\frac{1}{8\pi}\int_{\partial\mathcal{M}}\!\!d^3x\sqrt{k}\,\big(K-K_0\big)
\end{equation}
$R$ is the Ricci scalar, $F$ is the field strength tensor, $K$ is the trace of the extrinsic curvature of the boundary, and $\Lambda$ is the cosmological constant, which is positive for asymptotically de-Sitter space. $K_0$ is a subtraction term which is chosen to normalize the action.
\\

We shall choose $K_0$ such that $I=0$ when the mass (and therefore charge) of the black hole goes to zero. In this way, we are using empty de Sitter space as the reference point from which energy is measured, as opposed to the topologically distinct Minkowski space.  This is in  contrast to the work of Carlip and Vaidya \cite{CarlipVaidya2003}, where the 
latter choice for boundary term $K_0$ was chosen, so that the action vanishes for flat space.  Each choice 
results in different values for most thermodynamic quantities and changes the location of the critical points, but we find that they produce the same qualitative behaviour. We will refer to the flat background choice for $K_0$ where appropriate.
\\

We take a spherically symmetric ansatz for the metric,
\begin{equation}
ds^2=f(y)^2d\tau^2+\alpha(y)^2dy^2+r(y)^2d\Omega^2
\end{equation}
where $y\in[0,1]$ is a compactified radial coordinate with $y=0$ corresponding to the black hole horizon ($r(0)=r_+$) and $y=1$ corresponding to the cavity wall ($r(1)=R_c$). The boundary at $y=1$ has topology $S^1\!\times\!S^2$ with $S^2$ having area $4\pi R_c^2$. Heat flux through the cavity wall is chosen such that its temperature $T=\beta^{-1}$ remains fixed. The inverse temperature $\beta$ is related to the proper length of the boundary $S^1$ by $\beta=2\pi f(1)$, where the periodicity in imaginary time $\tau$ is $2\pi$.
\\

Thermodynamic quantities can be derived from the action (2.2), after the integrations are carried out and the various constraint imposed to arrive at the reduced action $I_r$. This procedure is described in Appendix A. Extremizing the reduced action with respect to $r_+$ and solving for $\beta$ leads to an expression for the inverse temperature (and therefore the temperature $T$), which in general depends on the other physical degrees of freedom present:
\begin{equation}
\dfrac{\partial I_r(\beta,r_+,R_c,q,\Lambda)}{\partial r_+}=0\quad\rightarrow\quad\beta(r_+,R_c,q,\Lambda)\implies T(r_+,R_c,q,\Lambda)
\end{equation}

Taking derivatives of the reduced action with respect to the physical degrees of freedom at the stationary points gives the energy and entropy in this ensemble:
\begin{equation}
E=\dfrac{\partial I_r}{\partial \beta},\qquad S=\beta\left(\dfrac{\partial I_r}{\partial\beta}\right)-I_r
\end{equation}
The energy as defined is the mean thermal energy of the black hole with respect to the empty de Sitter spacetime, which can be related to the ADM mass of the spacetime after accounting for the gravitational binding energy and electrostatic binding energy\footnote{The thermal energy $E$ is related to the ADM mass $M$ of the spacetime through $M=E\sqrt{1-\frac{\Lambda R_c^2}{3}}-\frac{E^2}{2R_c}+\frac{q^2}{R_c^2}$, where $q$ is the charge and $R_c$ is the cavity radius.}.
\\

In the discussion of phase transitions, the thermodynamic potential of interest is the {\it Helmholtz free energy}, $F$. It is the potential that is minimized when a thermodynamic system reaches equilibrium at constant temperature. In the grand canonical ensemble, $F$ is defined by $F=E-TS$ and can therefore be deduced directly from (2.4) and (2.5). Plotting $F(T)$ for fixed pressure (and charge, if it is present) reveals whether any phase transitions occur in the system.

\subsection{The First Law}

The first law of thermodynamics for black holes, first formulated by Bardeen, Hawking, and Carter \cite{Bardeen1973} reads
\begin{equation}
dE=\frac{\kappa}{8\pi}dA+\Omega dJ+\phi dQ
\end{equation}
where $\kappa$ is the surface gravity of the black hole, $A$ is the horizon area, $J$ is the angular momentum, $\Omega$ is the angular velocity, $\phi$ is the electrostatic potential, and $Q$ is the electric charge. At the classical level this law is merely analogous to the ordinary first law of thermodynamics, since there is no notion of temperature or mechanism for equilibration in classical black holes. Once we move to the semi-classical regime, where black holes can radiate, this analogy is made precise by considering the thermodynamic temperature and entropy to be, respectively, the thermal energy of Hawking radiation and the number of horizon degrees of freedom.
\\

When studying the thermodynamics of asymptotically AdS black holes in the extended phase space, one usually identifies the mass $M$ of the black hole with the enthalpy $H=E+PV$, rather than the internal energy, $E$. In the case of de Sitter black holes in a cavity, we will find that in order for the first law (and Smarr relation) to be satisfied with the usual definitions of temperature, entropy, and pressure, we must identify the energy (2.5) with the internal energy of the thermodynamic system. We are thus led to the following form of the first law: 
\begin{equation}
dE=TdS-\lambda dA_c-VdP+\phi dQ
\end{equation}
The $\lambda dA_c$ term arises from the presence of the cavity, where $A_c=4\pi R_c^2$ is the cavity surface area and $\lambda$ is the surface tension/pressure, which can be positive or negative.
\\

The action allows us to directly determine the temperature $T$, entropy $S$, and energy $E$ of the system. 
%We have also identified $\Lambda=-8\pi P$ which determines the pressure. 
In order to establish the first law however, more information is required, namely the thermodynamic volume and surface tension. This is done by evaluating the differentials $dE$, $dS$, $dA_c$, and $dP$, then enforcing that (2.7) holds and solving for the unknowns $\lambda$ and $V$. The resulting thermodynamic volume will in general be different from the geometric volume $V=\tfrac{4}{3}\pi r_+^3$ (that appears for charged AdS black holes \cite{Kubiznak2012}) as we will see.

\subsection{Thermodynamics with $\Lambda$}

The first law (2.6) as originally formulated does not contain anything analogous to the pressure-volume term $PdV$ that appears in ordinary thermodynamics, since there is no notion of pressure or volume for a black hole. In recent years however, the idea that a cosmological constant can act as a thermodynamic pressure has been extensively studied (see for example \cite{Dolan2014}, \cite{Zou2014}, \cite{Ma2014}, \cite{Wei2014}). In this {\it extended phase space}, the cosmological constant acts like a pressure according to the identification
\begin{equation}
P=-\dfrac{\Lambda}{8\pi}
\end{equation}
with the quantity conjugate to $P$ being interpreted as the thermodynamic volume:
\begin{equation}
V=\left(\dfrac{\partial E}{\partial P}\right)_{S,T}
\end{equation}
The extended phase space thermodynamics of asymptotically anti-de Sitter spacetimes, where the negative cosmological constant corresponds to a positive pressure, has been explored extensively. As noted above, in de Sitter space the cosmological constant is positive, so the variable $P$ is better thought of as a {\it tension}. Including the pressure in the thermodynamic ensemble allows us to study the equation of state $P(T,v)$ of our system. Once the thermodynamic volume has been determined from the first law, the equation of state follows by inverting the temperature for $\Lambda$ and substituting into (2.8). The expression to be solved is generally quadratic in $\Lambda$, leading to two possible equations of state, though only one is physical in our case. 
\\

A van der Waals fluid is characterized by an equation of state 
\begin{equation}
P(T,v)=\dfrac{T}{v-b}-\dfrac{a}{v^2}
\end{equation} 
where $a$ accounts for the attractive forces between constituent particles and $b$ accounts for their finite size. This equation of state accurately models the behaviour of fluids above their critical temperature and also captures the behaviour of fluids at the liquid-gas transition once Maxwell's equal area law is included. This kind of equation of state arises in the study of many black hole systems \cite{Kubiznak2016}, where the transition between small and large black holes is qualitatively identical to the liquid-gas transition modeled by the van der Waals equation of state. We will see here a departure from this behaviour, where the presence of the isothermal cavity leads to an equation of state that is non-linear in $T$. 
\\

The inclusion of $\Lambda$ and the isothermal cavity in the thermodynamic description also leads to a modified Smarr relation which in four dimensions is 
\be
E=2(TS-\lambda A_c+PV)+\Phi Q
\ee
and can be derived from various scaling arguments \cite{Smarr1973}. In the absence of the isothermal cavity, the $\lambda dA_c$ 
term vanishes and we recover the usual form of the Smarr relation. This relation is broadly applicable as it holds for both asymptotically AdS and dS spacetimes, is valid in any dimension, and is also satisfied by more exotic objects like black rings and black branes \cite{Kastor2009}. 

\section{Schwarzschild-de Sitter Black Holes}

With the general formalism in place, we move to the discussion of the (uncharged) Schwarzschild-de Sitter black hole. Solving the Hamiltonian constraint leads to the metric function 
\begin{equation}
N(r)=1-\dfrac{2m}{r}-\dfrac{\Lambda r^2}{3}
\end{equation}
The horizons are located at the real and positive roots of $N(r)=0$, of which there are two for the parameter range $0<9\Lambda m^2<1$. The smaller root $r_+$ gives the location of the event horizon, while the larger root $r_{c}$ is the cosmological horizon. The limit where the horizons coincide is known as the Nariai limit \cite{Nariai1951}, where $9\Lambda m^2\rightarrow 1$ and thus $r_+\rightarrow R_c$.
\\

Following the methods in Appendix A, we can perform the integrations in (2.2) to arrive at the reduced action:
\begin{equation}
I=\beta R_c\left[\sqrt{1-\frac{\Lambda R_c^2}{3}}-\sqrt{\left(1-\frac{r_+}{R_c}\right)\left(1-\frac{\Lambda}{3}\big(R_c^2+R_cr_++r_+^2\big)\right)}\ \right]-\pi r_+^2
\end{equation}
The inverse temperature is found by extremizing the action with respect to $r_+$ and solving for $\beta$, giving
\begin{equation}
\beta=\dfrac{4\pi r_+\sqrt{\left(1-\frac{r_+}{R_c}\right)\Big(1-\frac{\Lambda}{3}\big(R_c^2+R_cr_++r_+^2\big)\Big)}}{1-\Lambda r_+^2}
\end{equation}
from which the temperature is
\begin{equation}\label{temp}
T=\dfrac{1-\Lambda r_+^2}{4\pi r_+\sqrt{\left(1-\frac{r_+}{R_c}\right)\Big(1-\frac{\Lambda}{3}\big(R_c^2+R_cr_++r_+^2\big)\Big)}}
\end{equation}
In the limit $\Lambda\rightarrow0,\ R_c\rightarrow\infty$, this reduces to the familiar result $T=1/4\pi r_+$. The entropy is
\begin{equation}
S=\beta\dfrac{\partial I}{\partial\beta}-I=\pi r_+^2
\end{equation}
and finally the energy is
\begin{equation}
E=\dfrac{\partial I}{\partial \beta}=R_c\left[\sqrt{1-\frac{\Lambda R_c^2}{3}}-\sqrt{\left(1-\frac{r_+}{R_c}\right)\left(1-\frac{\Lambda}{3}\big(R_c^2+R_cr_++r_+^2\big)\right)}\ \right]
\end{equation}

\subsection{The First Law}

With the energy $E$, temperature $T$, and entropy $S$ as defined above, we can determine what the thermodynamic volume $V$ and surface tension $\lambda$ must be for the first law (2.7) to hold. Equating the left and right hand sides of (2.7) requires that
\begin{equation}
\lambda=\dfrac{\left(4\Lambda R_c^3-6R_c\right)\left(X-Y\right)+r_+(\Lambda r_+^2-3) Y}{48 \pi R_c^2 XY}
\end{equation}
\begin{equation}\label{XYvol}
V=\dfrac{4\pi}{3}\,\dfrac{R_c^3\,(Y-X)-r_+^3 Y}{XY}
\end{equation}
where we have defined the quantities
\begin{equation}\label{XYdef}
X=X(\Lambda)\equiv\sqrt{\left(1-\frac{r_+}{R_c}\right)\bigg(1-\frac{\Lambda}{3}\big(R_c^2+R_cr_++r_+^2\big)\bigg)},\qquad Y=Y(\Lambda)\equiv\sqrt{1-\dfrac{\Lambda R_c^2}{3}}
\end{equation}
With these definitions it is straightforward to verify that  
\begin{equation}
dE=TdS-\lambda dA_c-VdP
\end{equation}
where the sign of the $VdP$ term is negative in order to ensure that the thermodynamic volume is positive in the appropriate region. Unlike $V$, the sign of $\lambda$ is free to be negative since this corresponds to a surface pressure that is still physical, unlike a negative volume. Figure 1 shows regions in parameter space where $V$ and $\lambda$ are positive, in terms of the dimensionless ratio $x\equiv r_+/R_c\in[0,1]$.
\\

\begin{figure}[h]
\includegraphics[width=0.45\textwidth]{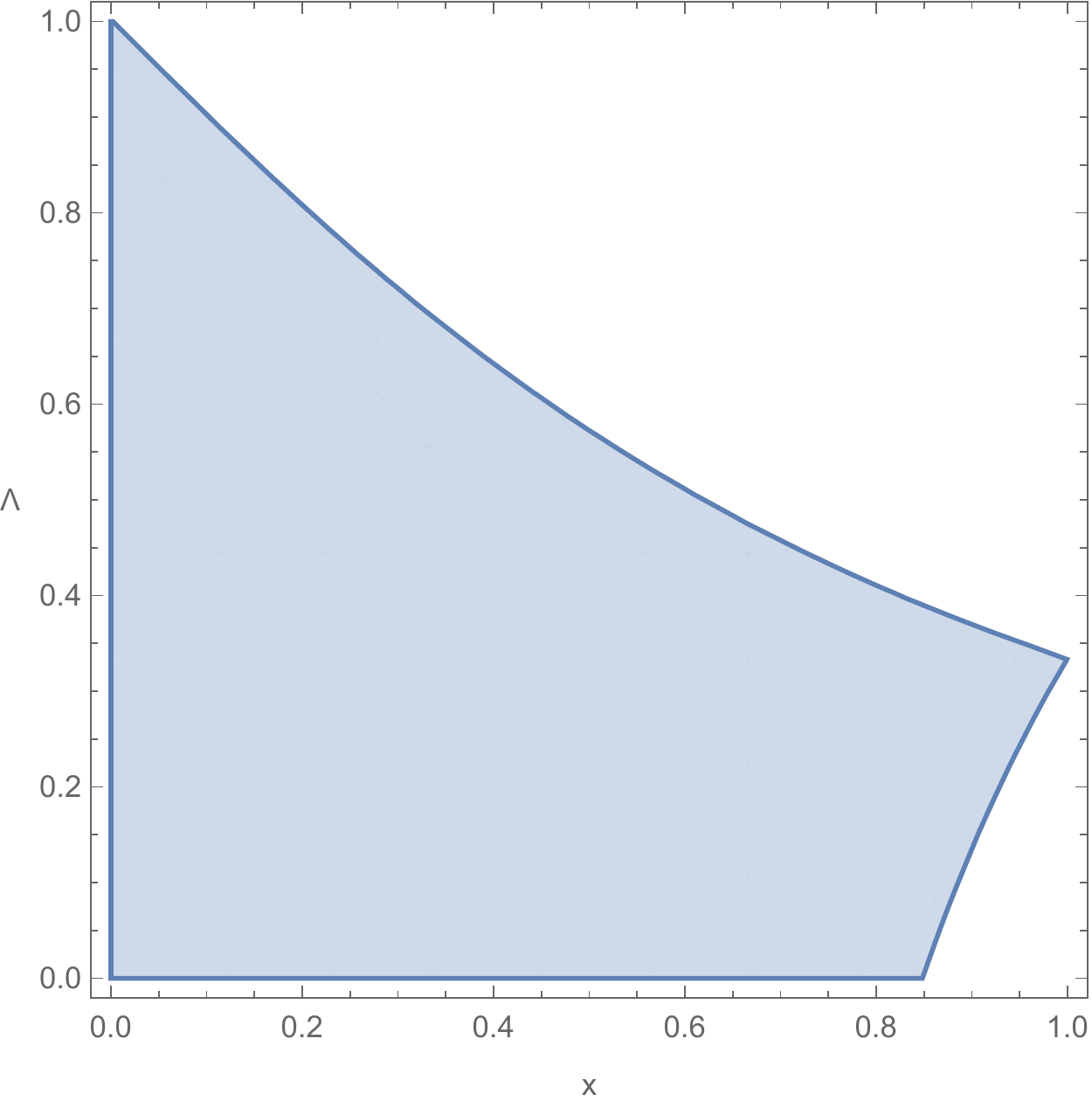}\qquad\includegraphics[width=0.45\textwidth]{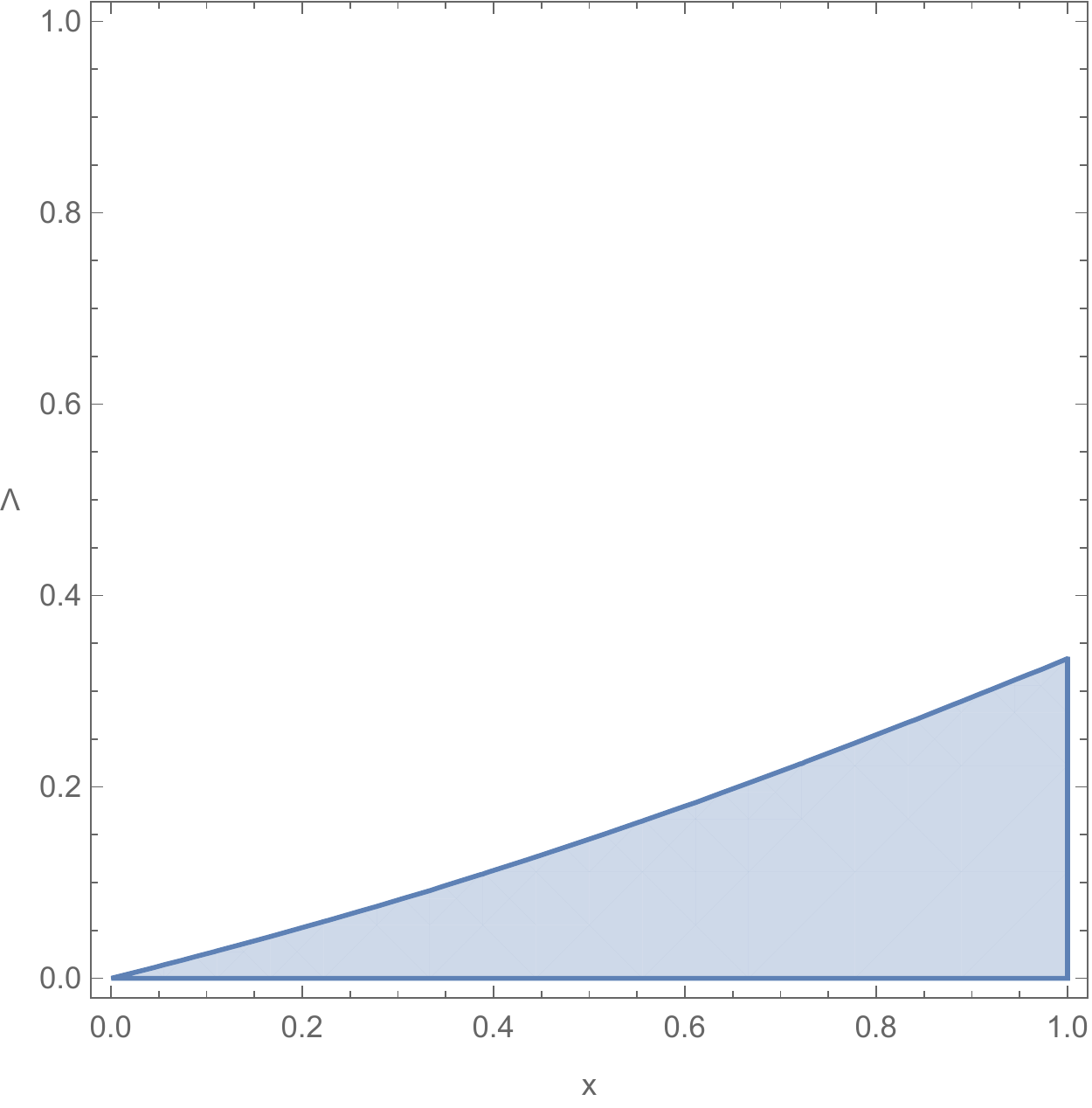}
\caption{Regions of positivity for the thermodynamic volume $V$ (left) and surface tension $\lambda$ (right) as a function of $x$ and $\Lambda$ for fixed cavity radius ($R_c=\sqrt{3}$) and varying charge. Blue shaded regions indicate where the respective quantity is positive.}
\end{figure}

There are two limits of interest here. The first is the asymptotically flat limit $\Lambda\rightarrow 0$. In this case we have $X\rightarrow \sqrt{1-r_+/R_c}$ and $Y\rightarrow 1$, giving
\begin{equation}\label{lamVol}
\lambda=\dfrac{2R_c\left(1-\sqrt{1-\dfrac{r_+}{R_c}}\right)-r_+}{16 \pi R_c^2 \sqrt{1-\dfrac{r_+}{R_c}}}, \qquad V=\dfrac{4\pi}{3}\,\dfrac{R_c^3\,\left(1-\sqrt{1-\dfrac{r_+}{R_c}}\right)-r_+^3}{\sqrt{1-\dfrac{r_+}{R_c}}}
\end{equation}
which agrees with the results of Brown et al \cite{Brown1990}. Note that in this limit, $V>0$ provided that $r_+<0.8483748957 R_c$, as demonstrated in Figure 1. Since there is no longer a cosmological horizon in this limit, we can further take the large cavity limit $R_c\rightarrow \infty$, which results in $\lambda\rightarrow 0$ and $V\rightarrow\infty$, as expected. 
\\

We can also take the small black hole limit $r_+\rightarrow 0$, where $\lambda\rightarrow 0$ and $V\rightarrow 0$.
%The next limit we examine is when the black hole is very small compared to the cavity size but the cosmological constant is non-zero. 
In this case we let $r_+=x\,R_c$ and expand around $x=0$ to find
\begin{equation}\label{Vapprox}
\lambda\approx-\dfrac{\sqrt{3}R_c\Lambda}{16\pi (3-\Lambda R_c^2)^{3/2}}\,x+\mathcal{O}(x^2)\ ,\qquad V\approx\dfrac{6\pi R_c^3 x}{(3-\Lambda R_c^2)\sqrt{9-3\Lambda R_c^2}}\left(1-\frac{9x}{4(3-\Lambda R_c^2)}\right)+\mathcal{O}(x^3)
\end{equation}
Both the thermodynamic volume and surface tension vanish in the small black hole limit for fixed cavity size. Of course, we cannot take the large cavity limit here because the cavity size is bounded by the cosmological horizon.

\subsection{Helmholtz Free Energy and Phase Transitions}

We can now examine the phase structure of the uncharged Schwarzschild-de Sitter black hole. To do this we look at the  Helmholtz free energy, $F=E-TS$, whose global minimum corresponds to the equilibrium state of the system. A plot of $F$ as a function of $T$ and $P$ for fixed $R_c$ will reveal the presence of any phase transitions in the system. Using (3.4), (3.5) and (3.6) we have
\begin{align}
F(r_+,R_c,P)&=\dfrac{\Lambda r_+^3}{A}-R_cX+\dfrac{R_c Y}{\sqrt{3}}-\dfrac{r_+}{4A}\\
T(r_+,R_c,P)&=\dfrac{1-\Lambda r_+^2}{4\pi r_+ A}
\end{align}
where $F$ and $T$ are understood to be functions of $P$ through the definitions (3.9)
of $X$ and $Y$ and the identification $\Lambda=-8\pi P$. We plot $F(T)$ parametrically for fixed $P$ and $R_c$ using $r_+$ as the parameter. This is shown in Figures 2 and 3, which reveal the presence of a standard Hawking-Page phase transition from pure radiation to a black hole, whose size increases with increasing temperature. The $F=0$ line corresponds to the radiation phase, with the transition to a black hole occurring at the temperature $T_c$ where the blue line crosses $F=0$. Above this critical temperature the black hole phase has lower free energy and is thermodynamically preferred. The value of $T_c$ is the solution to a fifth-degree polynomial and therefore must be found numerically. The kink in the $F-T$ curve corresponds to $T_{min}$, the lowest temperature at which a locally stable (supercooled) small black hole can exist at the given pressure and cavity size. This point occurs where $\partial F/\partial T$ becomes undefined, and can be found analytically. The expression for $T_{min}$ is long and without much insight, so we omit it.
\\

\begin{figure}[h]
\includegraphics[width=0.45\textwidth]{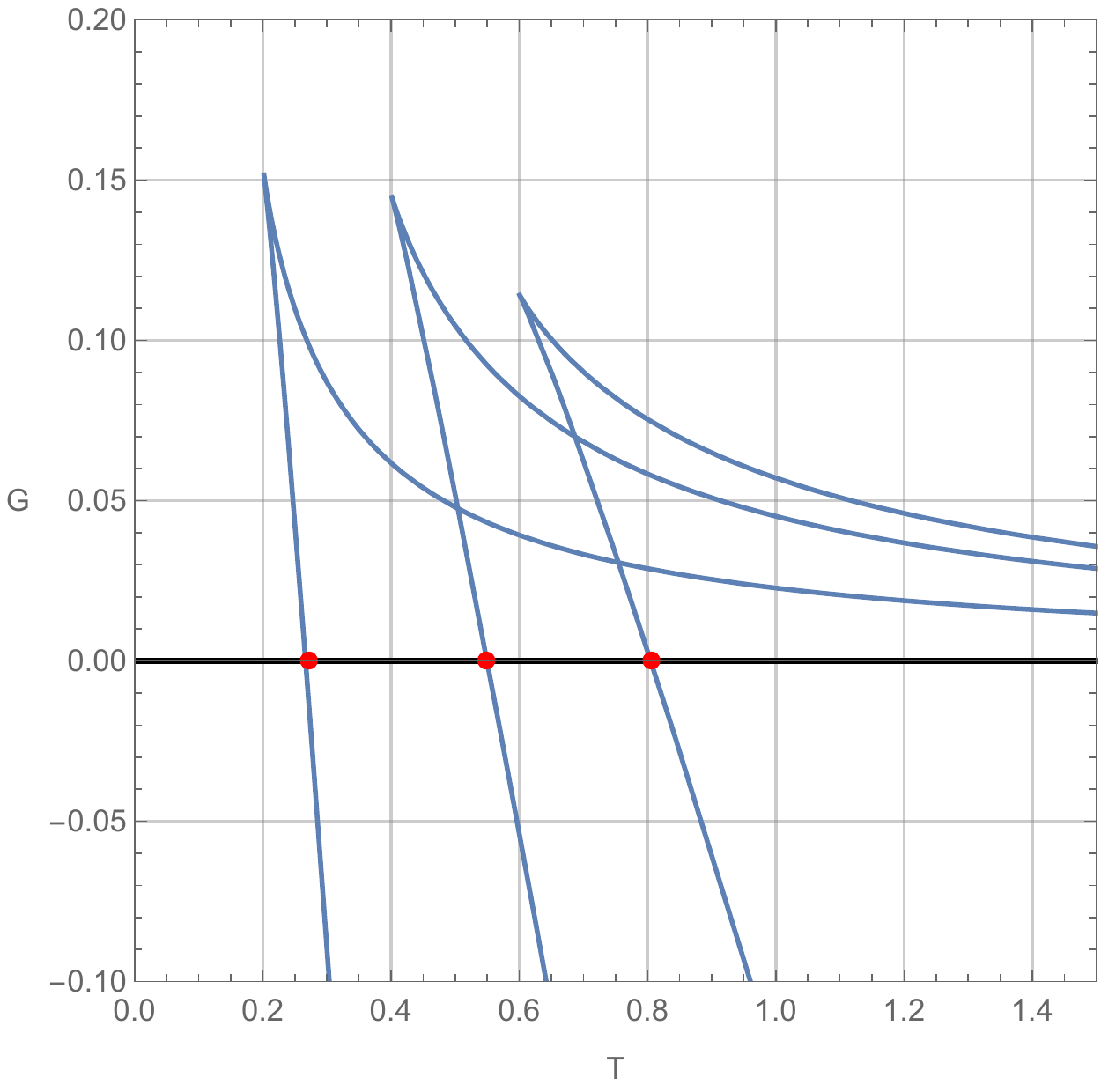}\qquad\includegraphics[width=0.45\textwidth]{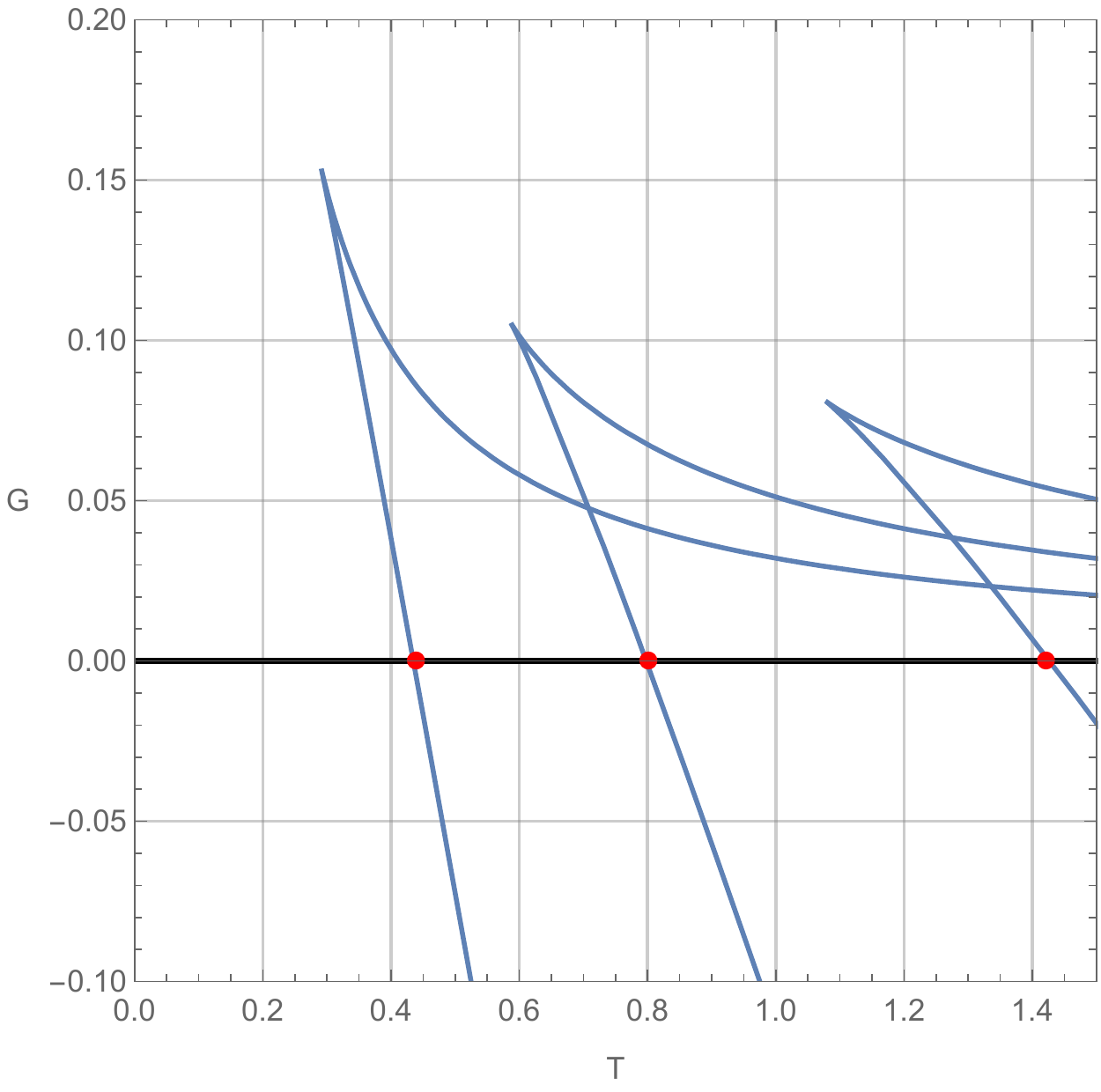}
\caption{Helmholtz free energy of the Schwarzschild-de Sitter black hole. \textbf{Left:} $F(T)$ for fixed cavity size ($R_c=1$) and varying pressure ($P=-0.01,-0.06,-0.07)$. \textbf{Right:} $F(T)$ for fixed pressure ($P=-0.1$) and varying cavity size ($R_c=0.6,0.8,0.9$). The critical temperature $T_c$ is indicated with a red dot.} 
\end{figure}

\begin{figure}[h]
\hspace{2.2cm}\includegraphics[width=0.7\textwidth]{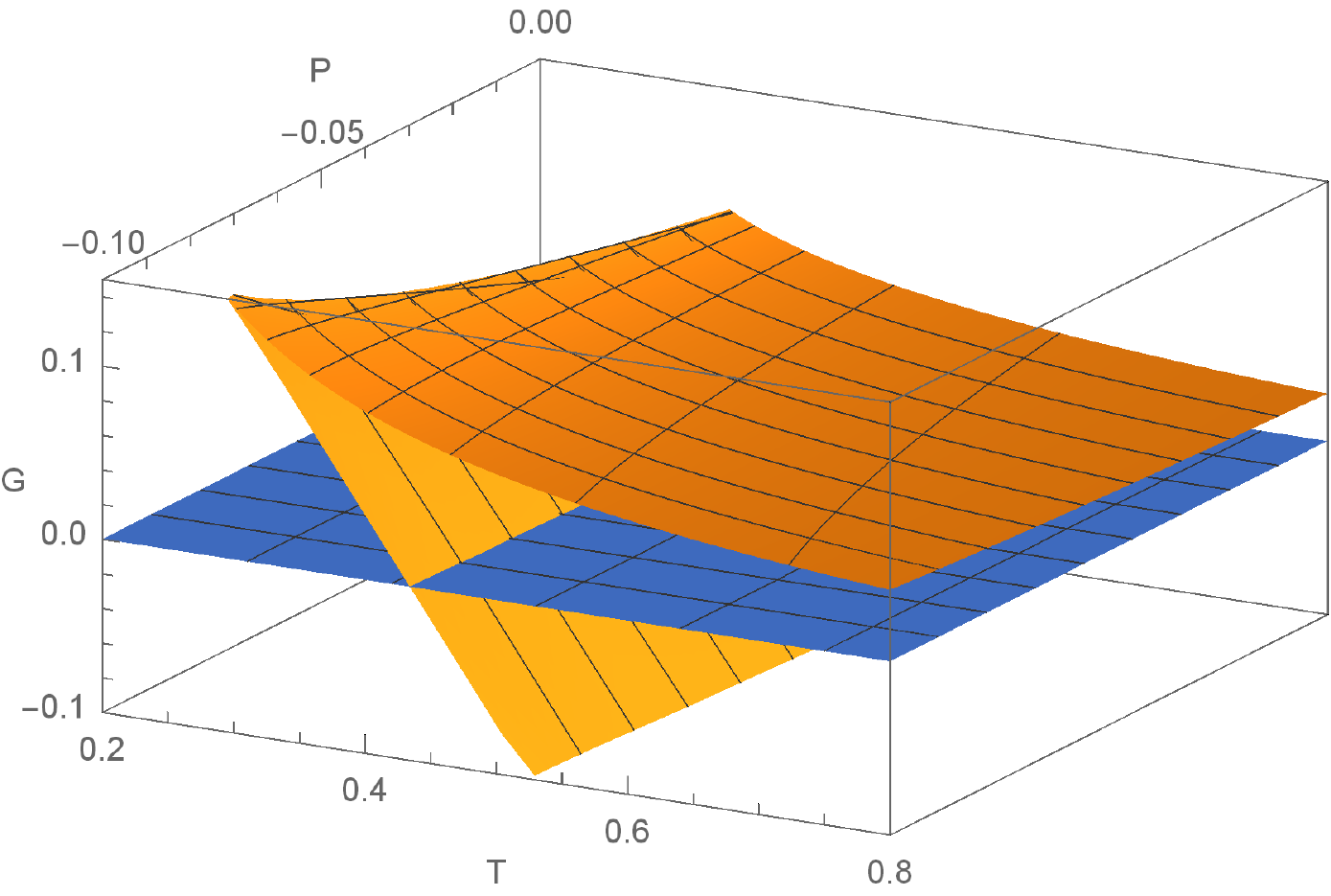}
\caption{Helmholtz free energy of the Schwarzschild-de Sitter black hole as a function of temperature and pressure for fixed charge ($q=0.15$). The $F=0$ plane is indicated in blue.}
\centering
\end{figure}

One can also verify that the heat capacity $C_v=-\beta(\partial S/\partial \beta)$ is positive above the transition temperature, demonstrating the stability of the black hole phase. Varying either the cosmological constant $\Lambda$ or the cavity size $R_c$ simply moves the location of the critical temperature, but does not qualitatively change the behaviour of the phase transition. This agrees with the results of Carlip and Vaidya \cite{CarlipVaidya2003}; the flat-background choice of normalization for the action simply moves the location of the critical point but does not introduce new thermodynamic features.
\\

As outlined in Section 2.3, working in the extended phase space allows us to examine the equation of state for uncharged de Sitter black holes. This is a relationship between the pressure $P$, temperature $T$, and volume $V$, which is the thermodynamic volume divided by the number of degrees of freedom associated with the horizon \cite{Kubiznak2016}.  In AdS space the volume $V\sim r_+^3$ and the specific volume (the volume per horizon degrees of freedom) $v \sim r_+$ \cite{Kubiznak2012}. However in dS space,  \eqref{temp} means that the pressure is a non-linear function of $(T,r_+)$, and \eqref{lamVol} in turn implies that $r_+$ is a highly non-linear function of $V$. As a result, the equation of state cannot be expressed in closed form. Implicitly plotting $P(V)$ at fixed $T$ reveals an absence of the oscillations characteristic of the van der Waals fluid. We omit the plot here for lack of insight.

\section{Reissner-Nordstrom-de Sitter Black Holes}

We now turn to the charged Reissner-Nordstrom-de Sitter black hole. The inclusion of charge alters the phase structure of black holes significantly, owing to the fact that the heat capacity can be positive and suffers from discontinuities \cite{Pavon1991}. In AdS, the presence of charge allows for small-to-large black hole phase transitions in the canonical ensemble \cite{Kubiznak2012}. Some of the thermodynamic properties of charged de Sitter black holes in a cavity have already been investigated by Carlip and Vaidya \cite{CarlipVaidya2003}. Here we explore their phase structure further, working in the extended phase space, and examine more closely the nature of the phase transitions present. 
\\

\noindent The metric function for the 4-dimensional charged de Sitter black hole is
\begin{equation}\label{chargedmetric}
N(r)=1-\dfrac{2m}{r}+\dfrac{q^2}{r^2}-\dfrac{\Lambda r^2}{3}
\end{equation}
and the resulting reduced action is
\begin{equation}
I=\beta R_c\left[\sqrt{1-\frac{\Lambda R_c^2}{3}}-\sqrt{\left(1-\frac{r_+}{R_c}\right)\left(1-\dfrac{q^2}{r_+R_c}-\frac{\Lambda}{3}\big(R_c^2+R_cr_++r_+^2\big)\right)}\ \right]-\pi r_+^2
\end{equation}
The expression \eqref{chargedmetric} for $r^2 N$ is now a quartic polynomial in $r$, and we must restrict ourselves to regions where $N(r)>0$ for the path integral to be well defined. An analysis of these allowed regions can be found in \cite{CarlipVaidya2003}.
\\

The temperature is found again by extremizing the action with respect to $r_+$ and solving for $\beta$, giving
\begin{equation}
T=\dfrac{1}{\beta}=\dfrac{1-\frac{q^2}{r_+R_c}-\frac{\Lambda}{3}\big(R_c^2+R_cr_++r_+^2\big)+\left(1-\frac{r_+}{R_c}\right)\left(\frac{\Lambda}{3}(R_c^2+2r_+R_c)-\frac{q^2}{r_+^2}\right)}{4\pi r_+\sqrt{\left(1-\frac{r_+}{R_c}\right)\Big(1-\frac{q^2}{r_+R_c}-\frac{\Lambda}{3}\big(R_c^2+R_cr_++r_+^2\big)\Big)}}.
\end{equation}
The entropy is again
\begin{equation}
S=\beta\dfrac{\partial I}{\partial\beta}-I=\pi r_+^2
\end{equation}
and the energy is
\begin{equation}
E=\dfrac{\partial I}{\partial \beta}=R_c\left[\sqrt{1-\frac{\Lambda R_c^2}{3}}-\sqrt{\left(1-\frac{r_+}{R_c}\right)\left(1-\dfrac{q^2}{r_+R_c}-\frac{\Lambda}{3}\big(R_c^2+R_cr_++r_+^2\big)\right)}\ \right].
\end{equation}

\subsection{The First Law}

As in Section 3, we enforce the first law to find the thermodynamic volume and surface tension/pressure. There is an additional term $\Phi dQ$ in the first law from the presence of charge. We find that 
\begin{equation}
\lambda=\dfrac{\left(4\Lambda R_c^3-6R_c\right)\big(\tilde{X}-\tilde{Y}\big)+r_+(\Lambda r_+^2-3) \tilde{Y}-\dfrac{3q^2 \tilde{Y}}{r_+}}{48 \pi R_c^2 \tilde{X}\tilde{Y}}
\end{equation}
\begin{equation}
V=\dfrac{4\pi}{3}\,\dfrac{R_c^3\,(\tilde{Y}-\tilde{X})-r_+^3 \tilde{Y}}{\tilde{X}\tilde{Y}},\qquad \Phi=\dfrac{(R_c-r_+)q}{r_+ R_c \tilde{X}}
\end{equation}
where we have now defined the quantities
\begin{equation}
\tilde{X}(\Lambda)\equiv\sqrt{\left(1-\frac{r_+}{R_c}\right)\bigg(1-\dfrac{q^2}{r_+ R_c}-\frac{\Lambda}{3}\big(R_c^2+R_cr_++r_+^2\big)\bigg)},\qquad \tilde{Y}(\Lambda)\equiv\sqrt{1-\dfrac{\Lambda R_c^2}{3}}
\end{equation}
With these definitions we have that both the first law, \begin{equation}
dE=TdS-\lambda dA_c-VdP+\Phi dQ.
\end{equation}
and Smarr relation,
\be
E=2(TS-\lambda A_c+PV)+\Phi Q
\ee
are satisfied. As before, the sign of the $VdP$ term is negative in order to ensure that the thermodynamic volume is positive. 
\\

We again depict the regions of positivity for $V$ and $\lambda$ in Figure \ref{chargedfig}. Note that as charge increases the region of positivity shrinks.

\begin{figure}[h]
\hspace{0.7cm}\includegraphics[width=0.45\textwidth]{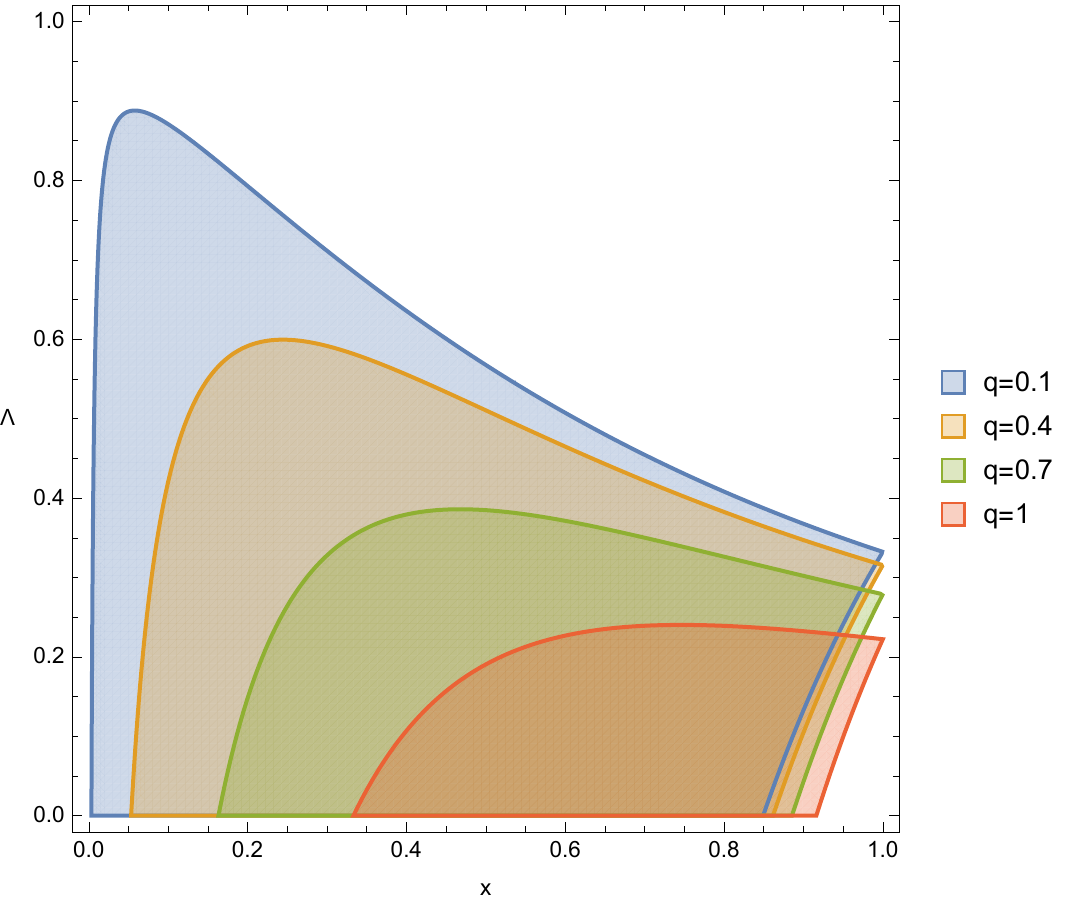}\qquad\includegraphics[width=0.45\textwidth]{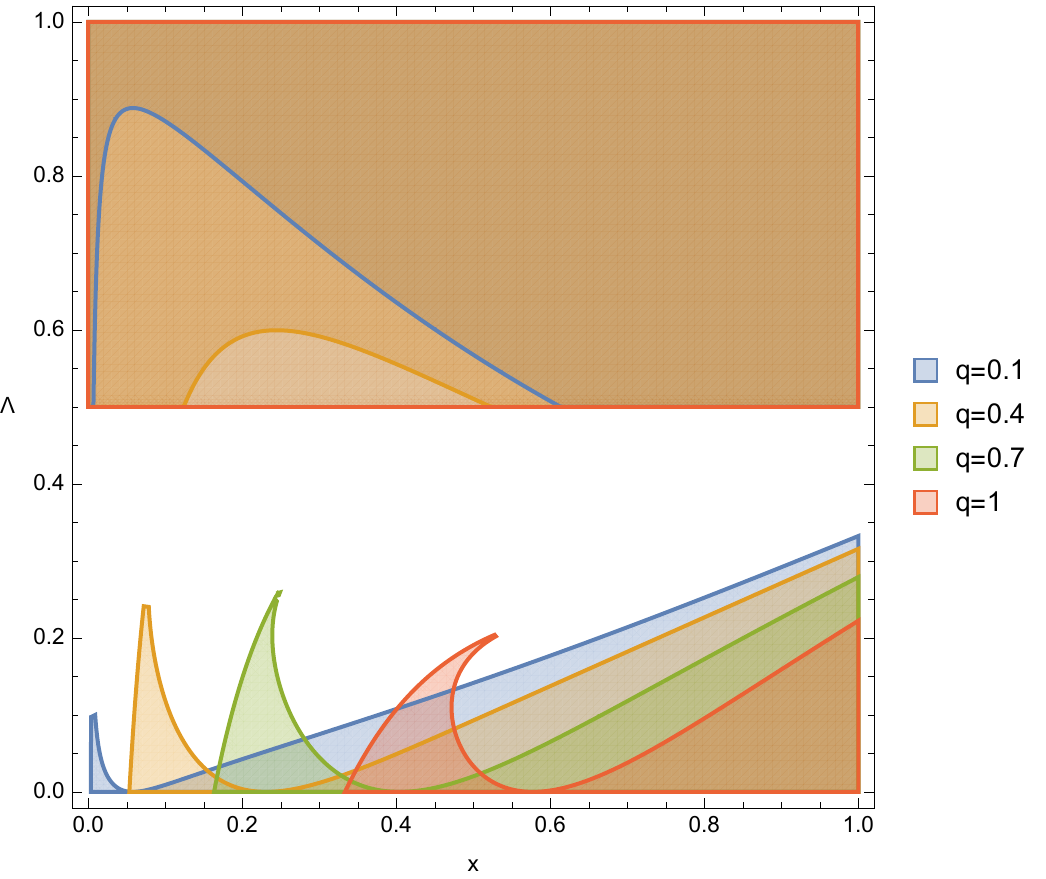}
\caption{Regions of positivity for the thermodynamic volume $V$ (left) and surface tension $\lambda$ (right) as a function of  $x= r_+/R_c$ and $\Lambda$ for fixed cavity radius $R_c=\sqrt{3}$. Shaded regions indicate where the respective quantity is positive.
}
\end{figure}

\subsection{Helmholtz Free Energy and Phase Transitions}

We turn once again to the Helmholtz free energy to look for phase transitions in the system. Using (4.3), (4.4) and (4.5) we have
\begin{align}
F(r_+,R_c,P)&=\dfrac{\Lambda r_+^3}{\tilde{X}}-R_c\tilde{X}+\dfrac{R_c B}{\sqrt{3}}-\dfrac{r_+}{4\tilde{X}}\\
T(r_+,R_c,P)&=\dfrac{1-\Lambda r_+^2}{4\pi r_+ \tilde{X}}
\end{align}
which is identical to the uncharged case aside from the extra $q$-dependent term appearing in $\tilde{X}$. We plot $F(T)$ parametrically for fixed $P$ and $R_c$ using $r_+$ as the parameter. This is shown in Figure 6.

\begin{figure}[h]
	\includegraphics[width=0.45\textwidth]{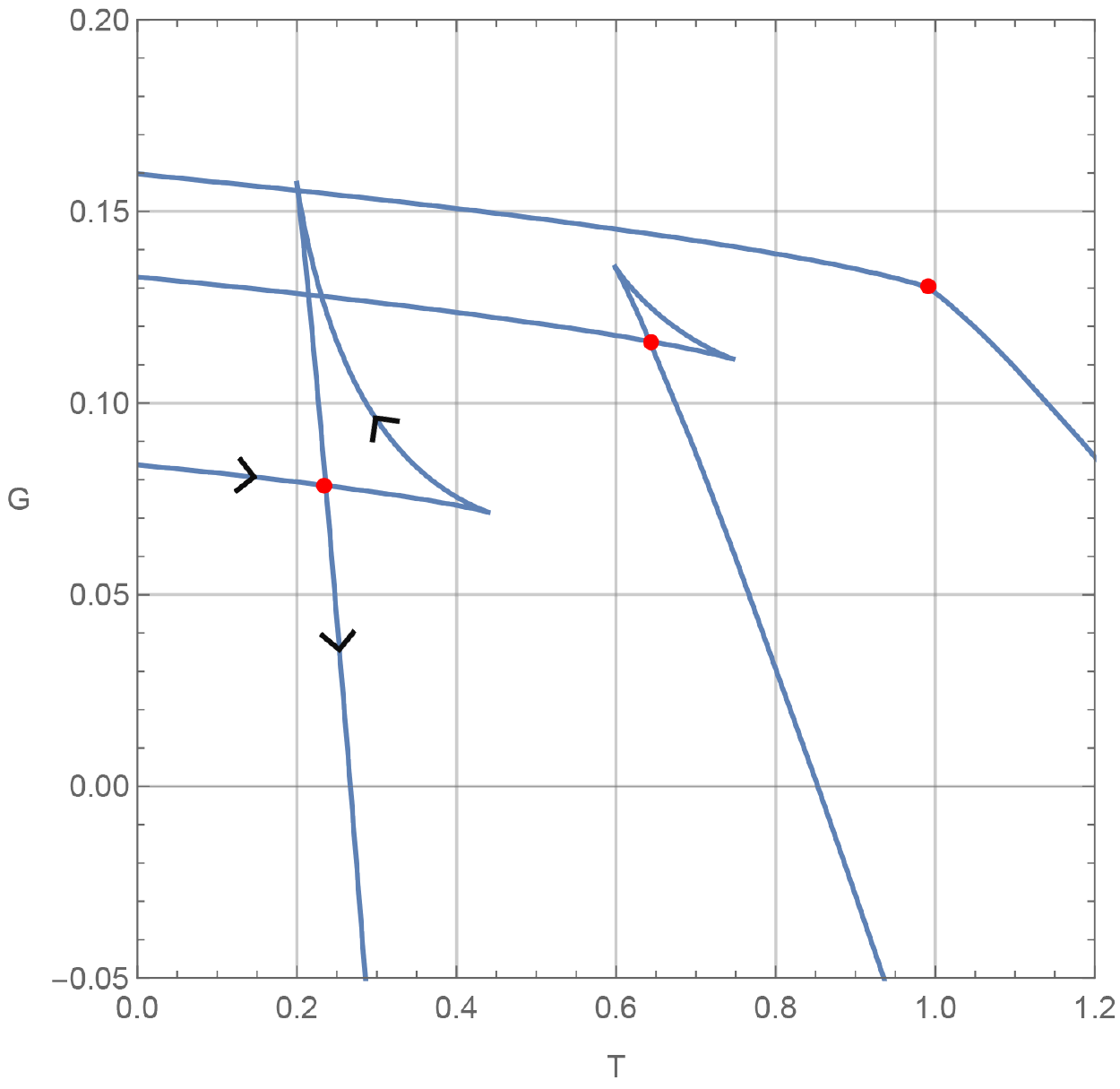}\qquad\includegraphics[width=0.45\textwidth]{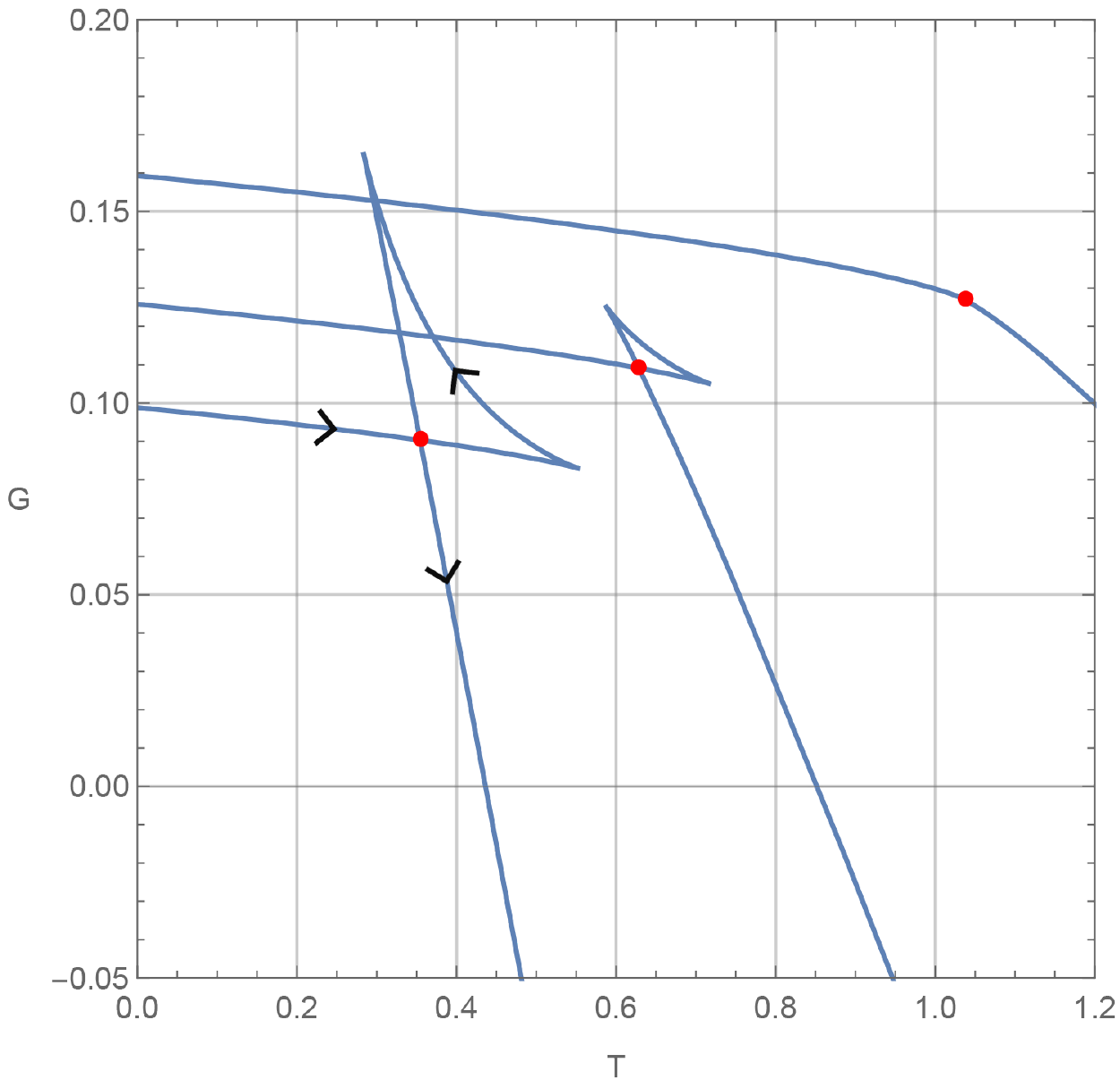}
	\caption{Helmholtz free energy of the Reissner-Nordstrom-de Sitter black hole as a function of temperature. {\bf Left:} Various pressures ($P=-0.082, -0.07, -0.01$) with fixed charge ($q=0.08$) and cavity size ($R_c=1$), showing clearly the formation of a swallowtail below the critical pressure, in this case $P_c\approx-0.82$. {\bf Right:} Various cavity sizes ($R_c=0.6,0.8,0.9$) with fixed charge charge ($q=0.08$) and pressure ($P=-0.08$). The red dots indicate the location of the small-to-large black hole phase transition.}
	\centering
\label{chargedfig}	
\end{figure}

Though the free energy is largely similar to that of (3.13), the presence of charge significantly alters the phase structure. Below a certain critical pressure $P_c$, the free energy develops a kink which in the $F-T-P$ space forms a swallowtail. The swallowtail indicates the presence of a first order phase transition from a small black hole to a large black hole, which occurs at the crossing where the critical temperature $T_c$ is reached.  The horizon radius $r_+$ increases along the near-horizontal line at the left (in the direction indicated by the arrow), which is
the same as the direction of increasing temperature.  Eventually  a crossover point (where $T=T_c$) is reached, beyond which the free energy is  minimized by moving downward along the steeper line instead of forward along the near-horizontal line. At $T=T_c$ there is a discontinuity in the horizon radius $r_+$.
\\

This swallowtail behaviour appears also in charged and rotating AdS black holes \cite{Simovic2014}, though we note a significant departure from the behaviour of previously studied examples in our case. Normally, as the magnitude of the pressure increases, the swallowtail grows without bound, indicating the presence of a phase transition for all pressures above the critical pressure $P_c$ where the swallowtail first forms. In our case however, the swallowtail closes past a certain maximum pressure $P_{max}$, creating a {\it swallowtube}. The phase transition from a small to large black hole is then {\it compact} in the sense that it exists only in a finite domain in $P$. For sufficiently large $R_c$ or small $q$, the swallowtube intersects the $P=0$ plane and is cut off (since we are in de Sitter space, $P$ must be negative), though it still closes off at $P_{max}$. This is qualitatively different from the reentrant phase transitions found in, for example, higher dimensional rotating AdS black holes, where lowering the temperature of the black hole results in a small-large-small phase transition \cite{Altamirano2013}. 
\\

It is interesting to examine the coexistence curve for this system, which we show in Figure 7. Here also we see a departure from the behaviour of AdS black holes, where the coexistence line increases monotonically and terminates at one end at a second order phase transition. Instead, the compact region that comprises the swallowtube is represented by a line segment in $P-T$ space along which the small and large black hole phase coexist, terminating at either end with a second order phase transition at the two critical pressures where the two ends of the swallowtube first form.

\begin{figure}
\centering
\includegraphics[width=0.6\textwidth]{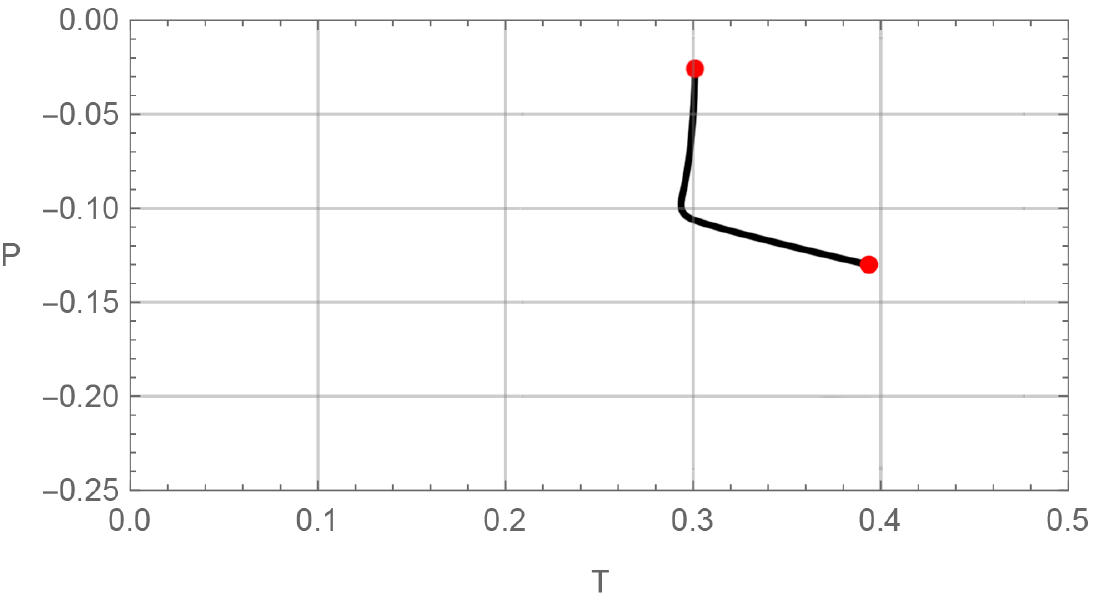}
	\caption{Coexistence line for the charged dS black hole, along which the small and large black hole phases coexist. The line terminates at two ends where a second order phase transition occurs, indicated by a red dot.}
	
\end{figure}

\begin{figure}[h]
	\hspace{1.8cm}\includegraphics[width=0.8\textwidth]{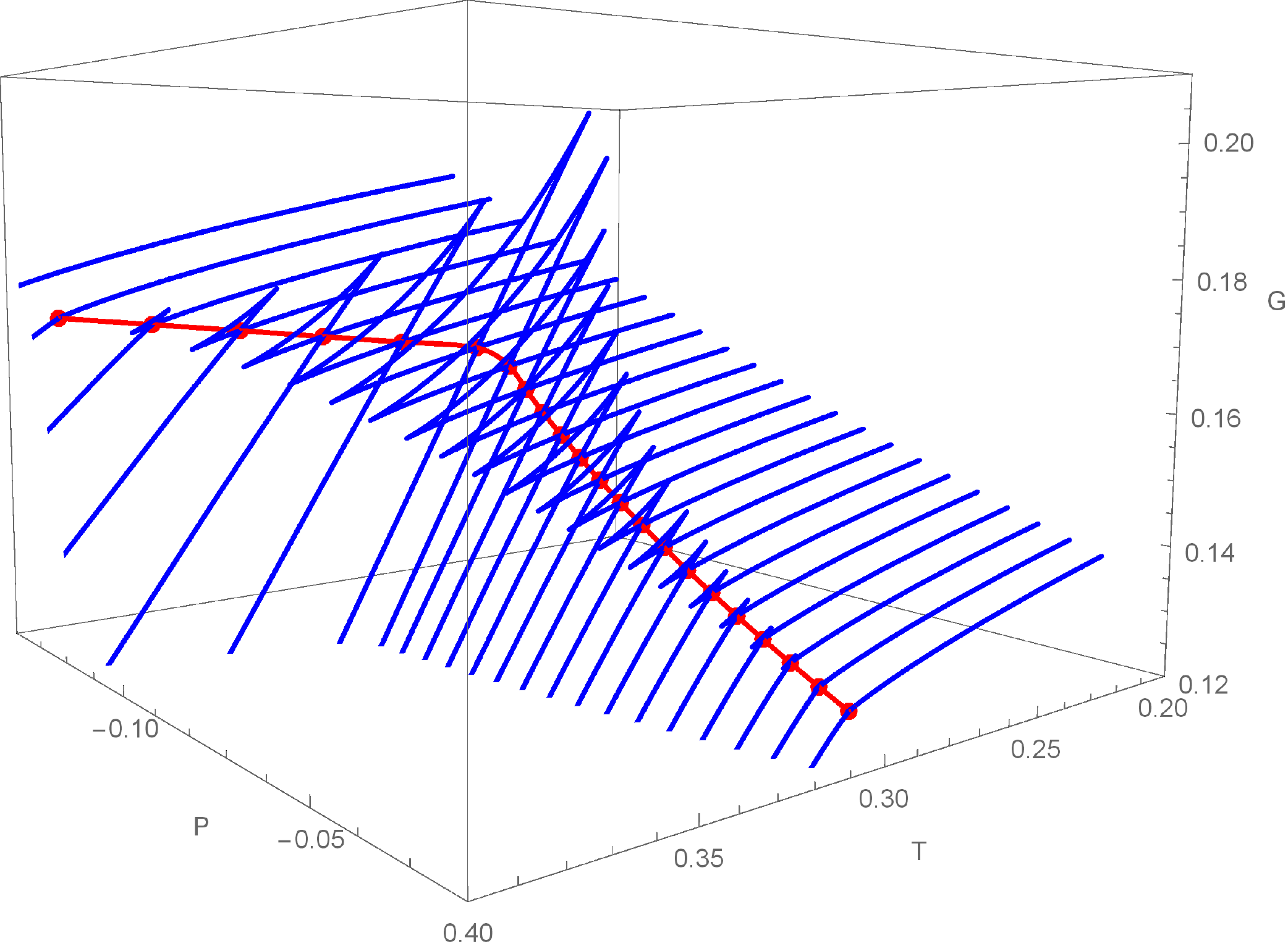}
	\caption{Helmholtz free energy of the Reissner-Nordstrom-de Sitter black hole as a function of temperature and pressure for fixed charge ($q=0.08$), showing slices of constant pressure, and demonstrating the compact nature of the phase transition.}
	\centering
\end{figure}
 
 To close this section, we again consider the equation of state of the system. Finding the pressure $P(T,V)$ by eliminating $r_+$ now requires solving $T=T(P,r_+,R_c,q)$, which is a ninth-degree polynomial in $r_+$. Again, one must proceed by plotting $P$ and $T$ implicitly using $r_+$ as the parameter. As in the uncharged case, implicitly plotting $P(V)$ for fixed $T$ reveals an absence of the oscillations characteristic of the van der Waals fluid.

\section{Conclusions}

The extended phase space thermodynamics of de Sitter black holes in an isothermal cavity is 
simultaneously quite similar and strikingly different from that of asymptotically flat and AdS black holes. 
Of immediate note is the fact that there is no simple equation of state relating the pressure to the thermodynamic volume. This is because in both the uncharged and charged cases the presence of a cavity necessarily introduces a complicated non-linear relationship between these quantities, and the presence of standard van der Waals
oscillations is no longer present.  Despite this, the free energy indicates clear and interesting phase behaviour
for both cases.
\\

For the uncharged case, the identification of the cosmological constant with the thermodynamic pressure does not change the phase structure considerably; we still note a first-order Hawking-Page phase transition from hot gas to a black hole, with a locally stable supercooled region. This is not surprising since the extended phase space Helmholtz free energy is identical to the `regular' phase space free energy. We also considered the analogy that these phase transitions make with van der Waals systems, and concluded that the equation of state does not support any liquid-gas type phase transition.  Distinct from the usual AdS case however is the fact that that the equation of state is highly non-linear in $T$. 
 \\
 
The charged de Sitter black hole presents novel features. We find a small-to-large black hole phase transition which occurs for pressures more negative than a certain critical pressure and above a certain critical charge. This structure is similar to the type seen in asymptotically AdS black holes. However this phase transition is compact in the sense that there is a second, more negative critical pressure $P_{max}$ below which the phase transition disappears. This is in stark contrast to what occurs in AdS black holes, where above the (positive) critical pressure the phase transition is always absent, and below this pressure the phase
transition is always present. For a large enough choice of cavity size, the phase transition is present even when $P\rightarrow 0$, but there is always a maximum pressure $|P_{max}|$ beyond which the small-large transition disappears, regardless of the choice of $q$ and $R_c$.
\\

There are many avenues to explore beyond the work presented here, the addition of angular momentum and working in higher dimensions being logical next steps. It will be particularly interesting to see if the reentrant phase transitions and triple points typical of higher dimensional AdS black holes manifest themselves in the presence of an isothermal cavity for other types of de Sitter black holes.

\section{Acknowledgements}This work was supported in part by the Natural Sciences and Engineering Council of Canada.

\clearpage

\appendix

\section{The Reduced Action}

Here we demonstrate the evaluation of the reduced action $I_r$ starting from the full action 
\begin{equation}
I=-\frac{1}{16\pi}\int_{\mathcal{M}}\!\!d^4x\sqrt{g}\,\big(R-2\Lambda+F^2\big)+\frac{1}{8\pi}\int_{\partial\mathcal{M}}\!\!d^3x\sqrt{k}\,\big(K-K_0\big)
\end{equation}
following the method of Brown \cite{Brown1990}. We begin with the following ansatz for the metric in terms of the radial coordinate $y\in[0,1]$:
\begin{equation}
ds^2=f(y)^2d\tau^2+\alpha(y)^2dy^2+r(y)^2d\Omega^2
\end{equation}
The horizon is located at $r_+=r(0)$, and has topology $\mathcal{S}^2$. The boundary has topology $S^1\times S^2$, is located at $y=1$, and has an $\mathcal{S}^2$ component (the cavity) with area $4\pi R_c^2$, where $R_c=r(1)$. The temperature of the boundary is given by the proper length of the circle $S^1$:
\begin{equation}
T^{-1}=\beta=\int_0^{2\pi}f(1)d\tau=2\pi f(1)
\end{equation}
The requirement that the horizon be $\mathcal{S}^2$ immediately implies that $f(0)=0$. This requires further that
\begin{equation}
\dfrac{f'}{\alpha}\bigg|_{y=0}=0
\end{equation}
in order for the near-horizon geometry to be flat. With the constraints in place we first examine the bulk part of the action:
\begin{equation}
I_\mathcal{M}=-\frac{1}{16\pi}\int_{\mathcal{M}}\!\!d^4x\,\sqrt{g}\,\big(R-2\Lambda+F^2\big)
\end{equation}
The Ricci scalar can be evaluated for the metric (A.2), giving
\begin{equation}
R=\frac{2f'\alpha'}{f\alpha^3}-\frac{2f''}{f\alpha^2}-\frac{4r'f'}{rf\alpha^2}+\frac{4r'\alpha'}{r\alpha^3}+\frac{2}{r^2}-\frac{2(r')^2}{r^2\alpha^2}
\end{equation}
where primes indicate derivatives with respect to $y$. The field strength tensor is $F_{\mu\nu}=\partial_{\mu}A_{\nu}-\partial_{\nu}A_{\mu}$. For a static, spherically symmetric spacetime, the gauge freedom allows us to write $A_{\mu}=A_{\tau}(y)$ with all other components vanishing. The boundedness of $A_{\mu}$ at $y=0$ and $y=1$ requires
\begin{equation}
A_{\tau}(0)=0,\qquad A_{\tau}(1)=\dfrac{k\beta \phi}{2\pi}
\end{equation}
where $\phi\equiv\phi(0)-\phi(1)$ and $k$ is a constant. With this, the Maxwell part of the action reduces to:
\begin{equation}
F^2=F_{\mu\nu}F^{\mu\nu}=\dfrac{2(A')^2}{\alpha^2 f^2}
\end{equation}
Substituting (A.6) and (A.8) into (A.5) and performing the integrations over $\theta$ and $\phi$ gives:
\begin{equation}
I_\mathcal{M}=\int\! dy\,d\tau\left[\dfrac{r^2f''}{2\alpha}+\dfrac{rr'f'}{\alpha}+\dfrac{rr''f}{\alpha}-\dfrac{\alpha f}{2}+\dfrac{(r')^2f}{\alpha}+\dfrac{r^2\alpha f\Lambda}{2}-\dfrac{r^2(A')^2}{2\alpha f}\right]
\end{equation}
Next we solve the Hamiltonian constraint $G{^{\tau}}_{\tau}+\Lambda g{^{\tau}}_{\tau}-8\pi T{^{\tau}}_{\tau}=0$ for $\Lambda$ and insert it into (A.9). The electromagnetic contributions cancel and we are left with:
\begin{equation}
I_\mathcal{M}=\int\! dy\,d\tau\left[\dfrac{r^2f''}{2\alpha}+\dfrac{rr'f'}{\alpha}+\dfrac{rr'f\alpha'}{\alpha^2}\right]
\end{equation}
Now we turn to the boundary action:
\begin{equation}
I_{\partial \mathcal{M}}=\frac{1}{8\pi}\int_{\partial\mathcal{M}}\!\!d^3x\,\sqrt{k}\,\big(K-K_0\big)
\end{equation}
$K$ is the trace of the extrinsic curvature of the boundary surface, $k$ is the metric on that surface, and the subtraction term $K_0$ is chosen such that $I=0$ when $m=0$. $K$ is defined in terms of the radial spacelike unit normal vector to the boundary $s^\mu=s^y=\alpha^{-1}(y)$ which gives
\begin{equation}
K=K_{\mu\nu}K^{\mu\nu}=-\dfrac{rf'+2r'f}{r\alpha f}
\end{equation}
where $K_{\mu\nu}=k{_{\mu}}^{\sigma}k{_{\nu}}^{\rho}\nabla_{\sigma}s_{\rho}$ and $k_{\mu\nu}=g_{\mu\nu}-s_{\mu}s_{\nu}$ is the metric on the boundary. Using (A.12) in (A.11) and integrating over $\phi$ and $\theta$ gives:
\begin{equation}
I_{\partial \mathcal{M}}=-\int\!d\tau\left[\dfrac{r^2f'}{2\alpha}+\dfrac{rr'f}{\alpha}+\dfrac{r^2fK_0}{2}\right]\,\bigg|_{y=1}
\end{equation}
\\
The full action is therefore:
\begin{equation}
I=\int\!dy\,d\tau\left[\dfrac{r^2f''}{2\alpha}+\dfrac{rr'f'}{\alpha}+\dfrac{rr'f\alpha'}{\alpha^2}\right]-\int\!d\tau\left[\dfrac{r^2f'}{2\alpha}+\dfrac{rr'f}{\alpha}+\dfrac{r^2fK_0}{2}\right]\,\bigg|_{y=1}
\end{equation}
Before proceeding further, we return to the Hamiltonian constraint, rewriting it as a total derivative with respect to $y$:
\begin{align}
G{^{\tau}}_{\tau}+\Lambda g{^{\tau}}_{\tau}-8\pi T{^\tau}_{\tau}=0&=\frac{(r')^2}{r^2\alpha^2}-\frac{1}{r^2}+\frac{2r''}{r\alpha^2}-\frac{2r'\alpha'}{r\alpha^3}+\Lambda+\dfrac{q^2}{r^4}\qquad\qquad\nonumber\\
0&=\frac{\left(\frac{r'}{\alpha}\right)^2-1}{r^2}+\frac{2}{r\alpha}\left(\frac{r''\alpha-r'\alpha'}{\alpha^2}\right)+\Lambda+\dfrac{q^2}{r^4}\nonumber\\
0&=\frac{1}{r^2r'}\left[r\left(\left(\frac{r'}{\alpha}\right)^2-1\right)\right]'+\Lambda+\dfrac{q^2}{r^4}\nonumber\\
0&=\left[r\left(\left(\frac{r'}{\alpha}\right)^2-1\right)\right]'+\frac{\Lambda}{3}[r^3]'-q^2\left[\dfrac{1}{r}\right]'
\end{align}
This is then integrated to obtain
\begin{equation}
\left(\frac{r'}{\alpha}\right)^2=1+\frac{C}{r}+\dfrac{q^2}{r^2}-\frac{\Lambda r^2}{3}
\end{equation}
The integration constant is found by requiring that the near-horizon geometry (the $y-\tau$ plane near $y=0$) be isometric to $\mathcal{R}^2$, and that the Euler number be $\chi=2$. With the given metric, these conditions imply that
\begin{equation}
\frac{f'}{\alpha}\,\bigg|_{y=0}=1\ ,\qquad \left(\frac{r'}{\alpha}\right)^{\!\!2}_{\!\!y=0}=0
\end{equation}
The second condition, along with (A.9) and the fact that $r(0)=r_+$ gives that
\begin{equation}
C=-r_+-\dfrac{q^2}{r_+}+\frac{\Lambda r_+^3}{3}
\end{equation}
which leads to
\begin{align}
\left(\frac{r'}{\alpha}\right)^{\!\!2}&=1-\dfrac{r_+}{r}-\dfrac{q^2}{r_+r}+\dfrac{\Lambda r_+^3}{3r}+\dfrac{q^2}{r^2}-\dfrac{\Lambda r^2}{3}\nonumber\\
&=\bigg(1-\frac{r_+}{r}\bigg)\!\left(1-\frac{q^2}{r_+r}-\frac{\Lambda}{3}\big(r^2+rr_++r_+^2\big)\right)
\end{align}
We now use the conditions (A.17) along with integration by parts and the product rule multiple times to rewrite the action (A.14) as a total derivative in $y$:
\begin{equation}
I=\int\!dy\,d\tau\left[\left(\dfrac{K_0\, r^2f}{2}\right)'-\left(\dfrac{rr'f}{\alpha}\right)'\right]=\int_0^{2\pi}\!d\tau\left[\dfrac{K_0\, r^2f}{2}-\dfrac{rr'f}{\alpha}\right]\,\bigg|_{y=0}^{y=1}
\end{equation}
Finally, we use the fact that $f(0)=0$, $r(1)=R_c$, $\beta=2\pi f(1)$, and (A.19) to arrive at:
\begin{align}
I&=\int_0^{2\pi}\!d\tau\left[\dfrac{K_0\, r^2f}{2}-\dfrac{rr'f}{\alpha}\right]\,\bigg|_{y=0}^{y=1}-\int_0^{2\pi}\!d\tau \left[\dfrac{r^2}{2}\right]\bigg|_{y=0}\nonumber\\
&=2\pi\left[\dfrac{K_0\, r(1)^2f(1)}{2}-r(1)f(1)\left(\dfrac{r'}{\alpha}\right)\right]-\pi r(0)^2\nonumber\\
&=\beta\, R_c\left[\dfrac{K_0 \,R_c}{2}-\sqrt{\bigg(1-\frac{r_+}{R_c}\bigg)\!\left(1-\frac{q^2}{r_+R_c}-\frac{\Lambda}{3}\big(R_c^2+R_cr_++r_+^2\big)\right)}\ \right]-\pi r_+^2
\end{align}
As noted in the main text, we choose $K_0$ such that the action vanishes when $m=0$ (which implies $q=0$ and $r_+=0$). This requires
\begin{equation}
K_0=\dfrac{2}{R_c}\sqrt{1-\dfrac{\Lambda R_c^2}{3}}
\end{equation}
which is easily verified to be the trace of the extrinsic curvature of the boundary when $m=0$. With this choice we arrive at the reduced action:
\begin{align}
I=\beta\, R_c\left[\sqrt{1-\dfrac{\Lambda R_c^2}{2}}-\sqrt{\bigg(1-\frac{r_+}{R_c}\bigg)\!\left(1-\frac{q^2}{r_+R_c}-\frac{\Lambda}{3}\big(R_c^2+R_cr_++r_+^2\big)\right)}\ \right]-\pi r_+^2
\end{align}


\begin{thebibliography}{30}

\bibitem{HawkingPage1983}
S. Hawking and D. Page, {\it Thermodynamics of black holes in anti-de Sitter space, Communications in Mathematical Physics}, 87, 577–588 (1983)

\bibitem{Maldacena1998}
J. M. Maldacena, {\it The Large N Limit of Superconformal Field Theories and Supergravity, International Journal of Theoretical Physics}, 38, 4, 1113-1133 (1999)

\bibitem{Policastro2002}
G. Policastro, D. T. Son, and A. O. Starinets, {\it From AdS/CFT correspondence to hydrodynamics, Journal of High Energy Physics}, 2002, 043 (2002)

\bibitem{Strominger2001}
A. Strominger, {\it The dS / CFT correspondence, Journal of High Energy Physics}, 0110, 034 (2001)

\bibitem{Sekiwa2006}
Y. Sekiwa, {\it Thermodynamics of de Sitter Black Holes: Thermal Cosmological Constant, Physical Review D}, 73, 084009 (2006)

\bibitem{Urano2009}
M. Urano, A. Tomimatsu, and H. Saida, {\it Mechanical First Law of Black Hole
Spacetimes with Cosmological Constant and Its Application to Schwarzschild-de Sitter
Spacetime, Classical and Quantum Gravity}, 26, 105010 (2009)

\bibitem{Brown1990} 
H. W. Braden, J. D. Brown, B. F. Whiting, J. W. York, {\it Charged black hole in a grand canonical ensemble, Physical Review D}, 42, 3376 (1990)

\bibitem{CarlipVaidya2003}
S. Carlip, S. Vaidya, {\it Phase Transitions and Critical Behavior for Charged Black Holes, Classical and Quantum Gravity}, 20, 3827-3838 (2003)

\bibitem{Kastor2009}
D. Kastor, S. Ray, J. Traschen, {\it Enthalpy and the Mechanics of AdS Black Holes, Classical and Quantum Gravity}, 26, 195011 (2009)

\bibitem{Kubiznak2016}
D. Kubiznak, R. B. Mann, M. Teo, {\it Black hole chemistry: thermodynamics with Lambda, Classical and Quantum Gravity}, 34, 063001 (2017)

\bibitem{Kubiznak2012}
D. Kubiznak, R. B. Mann, {\it P-V criticality of charged AdS black holes, Journal of High Energy Physics}, 1207, 033 (2012)

\bibitem{Parikh2002}
M. K. Parikh, {\it New Coordinates for de Sitter Space and de Sitter Radiation, Physics Letters B}, 546, 189-195 (2002)

\bibitem{Bardeen1973}
J. M. Bardeen, B. Carter, S. W. Hawking, {\it The four laws of black hole mechanics, Communications in Mathematical Physics}, 31, 161-170 (1973)

\bibitem{Smarr1973}
L. Smarr, {\it Mass formula for Kerr black holes, Physical Review Letters}, 30, 71-73 (1973)

\bibitem{Pavon1991}
D. Pavon, {\it Nonequilibrium fluctuations in cosmic vacuum decay, Physical Review D}, 43, 375 (1991)

\bibitem{Simovic2014}
A. M. Frassino, D. Kubiznak, R. B. Mann, F. Simovic, {\it Multiple Reentrant Phase Transitions and Triple Points in Lovelock Thermodynamics, Journal of High Energy Physics}, 8, (2014)



\bibitem{Altamirano2013}
N. Altamirano, D. Kubiznak, R. B. Mann, {\it Reentrant Phase Transitions in Rotating AdS Black Holes, Physical Review D}, 88, 101502 (2013)

\bibitem{Nariai1951}
H. Nariai, {\it On a new cosmological solution of Einstein's field equations of gravitation, The Science Reports of the Research Institutes, Tohoku University}, 35, 62 (1951)

\bibitem{Dolan2014}
B. P. Dolan, {\it The compressibility of rotating black holes in D-dimensions,
Classical and Quantum Gravity}, 31 035022 (2014)

\bibitem{Zou2014}
D.-C. Zou, S.-J. Zhang, B. Wang, {\it Critical behavior of Born-Infeld AdS black
holes in the extended phase space thermodynamics, Physical Review D}, 89, 044002 (2014)

\bibitem{Ma2014}
M. S. Ma, H. H. Zhao, L. C. Zhang, R. Zhao, {\it Existence condition and phase
transition of Reissner-Nordstrom-de Sitter black hole, International Journal of Modern Physics}, A29, 1450050 (2014)

\bibitem{Wei2014}
S. W. Wei, Y. X. Liu, {\it Triple points and phase diagrams in the extended phase
space of charged Gauss-Bonnet black holes in AdS space, Physical Review D}, 90, 044057 (2014)


\end{thebibliography}
\end{document}